# Driven localized excitations in the acoustic spectrum of small nonlinear macroscopic and microscopic lattices

M. Sato[†] and A. J. Sievers[‡]

† Graduate School of Natural Science and Technology, Kanazawa University

Kanazawa, Ishikawa 920-1192, Japan

‡Laboratory of Atomic and Solid State Physics, Cornell University

Ithaca, NY 14853-2501, USA

**Abstract**

Both bright and dark traveling, locked, intrinsic localized modes (ILMs) have been generated with a spatially uniform driver at a frequency in the acoustic spectrum of a nonlinear micromechanical cantilever array. Complementary numerical simulations show that a minimum density of modes, hence array size, is required for the formation of such locked smoothly running excitations. Additional simulations on a small 1-D antiferromagnetic spin system are used to illustrate that such uniformly driven running ILMs should be a generic feature of a nanoscale atomic lattice.

It has been predicted that a discrete lattice of the Fermi-Pasta-Ulam (FPU) type[1] in which the elements are coupled together nonlinearly would exibit intrinsic localized modes (ILMs)[2] as well as plane wave modes and such localized excitations have now been observed in a variety of discrete nonlinear systems [3]. Steady state driving[4,5] at the limiting frequencies of some dynamical systems have produced a systematic way to examine both the production and destruction of stationary ILMs[6-9]. Recent theoretical investigations have focused attention on the nonlinear dynamics of acoustic modes. In general dynamical energy cannot remain localized in the acoustic spectrum because some overtone of the ILM frequency is always on speaking terms with the linear phonon band. However, a theoretical study of finite-size Klein-Gordon lattices[10] has shown that for small systems where the phonon frequencies are sparsely distributed such localized solutions, referred to as "phantom breathers", can exist between these phonon modes. In a different direction theoretical studies on nonlinear plane wave excitations in FPU lattices, called q-breathers[11,12], are predicted to be stable for small enough nonlinearity and only require for their existence a discrete nonequidistant spectrum of normal modes, as induced by a finite system [13]. An experimental study of the acoustic nonlinear excitations for a small discrete FPU system is yet to occur.

This letter describes our experimental investigation of the dynamics of a small micromechanical cantilever array in the presence of a continuous, spacially uniform locking driver. With sufficient driving amplitude, beyond the q-breather limit, and with its frequency strategically located in the plane wave acoustic spectrum either bright or dark ILMs can be generated. We find that for a spatially uniform driver the running, locked ILM is a property of the small array. Using these findings as a starting point a small model atomic system in the form of a 1-D antiferromagnet is examined numerically to show that running, locked ILMs are a natural feature of a nanoscale lattice when excited by a spatially uniform driver.

The micromechanical di-element cantilever array, composed of long and short cantilevers, is made from a thin silicon nitride film (~300 nm thick) resting on a silicon substrate. Each cantilever has a transverse vibration mode and the coupling between cantilevers is provided by thin nitride overhang region. The number of cantilevers is 152 and the resonance frequencies are from 60 to 150 kHz. Both the cantilever itself and the interconnecting overhang have "hard" or "positive" nonlinearity. The sample is attached to a piezo-electric transducer (PZT) situated in a vacuum chamber. The array is shaken up and down uniformly. The vibration of each cantilever is recorded by a 1D-CCD camera. A laser beam is line focused along the cantilever tips, and the reflected beam is imaged on the camera using a lens. As the amplitude of a cantilever increases its image darkens. The array and measuring method are described in more detail in Ref. [14].

The observed localized vibrational patterns versus time are shown in Fig. 1. Panel (a) shows a stationary ILM driven at a frequency above the top of the plane wave spectrum. (Such stationary ILMs have been studied in detail in Ref. [14].) The horizontal white lines are images of cantilevers at rest. Because of the measuring technique the dark region near the center of the picture identifies large amplitude cantilevers associated with the ILM. The ILM, with locked amplitude, [4,8] is strongly pinned at a lattice site, is very stable and its step-wise manipulation has already been demonstrated [14]. Now examine the new finding in panel (b) in Fig. 1 where a driving frequency inside but near the top of the acoustic branch is used. A large amplitude (dark) region is generated and locked to the driver but now it travels across the array, reflecting from the boundaries, resulting in a zigzag pattern. Finally consider another new feature in panel (c) where a driving frequency is inside but near the bottom of the acoustic branch is used. The horizontal dark and white stripes represent a large amplitude standing wave which extends over all cantilevers. Superimposed on this pattern is a traveling dark ILM excitation. The arrows identify its zigzag path. The bright regions at the end points are due to overlapping incident and reflected hole excitations.

The experimental sequence to obtain a traveling ILM is as follows: identify a plane wave mode in the region where the nonlinearity balances the dispersion, slowly increase the driver frequency so the plane wave nonlinear mode grows in amplitude reaching the modulational instability, increase the driver frequency farther until a single traveling ILM is resolved. If no ILM state forms, repeat the sequence again from the beginning. The traveling ILM is one of these two stable states obtained by increasing the driving frequency beyond the unstable, modulation instability region, where many traveling wave packets are observed. When the driver frequency is decreased from either the ILM or the no-ILM high frequency state, the system always goes back into the unstable region. Only increasing the driver frequency provides the opportunity to reach the traveling ILM state. The traveling ILMs shown in Fig. 1 are obtained by increasing the driver frequency. For the 152 element array the spacing between plane wave modes with the same symmetry is about 1.5 kHz. When a locked traveling ILM appears its frequency can be varied over about ½ of that frequency interval. We now demonstrate that the precise behavior of these locked, and smoothly traveling ILMs is a feature of the small system size.

The acoustic and optic-like dispersion curves for the di-element array are shown in Fig. 2(a). Since the nonlinearity is positive a stationary locked ILM can appear above the upper branch. Although the highest and lowest frequency modes at $k = 0$ couple most strongly to the PZT, because of the fixed boundary conditions a single mode of odd symmetry in the lower branch (wavelength $\lambda = 2L/n$, where $L$ is the sample length and $n$ is an odd number) can be excited by simply adjusting the driver to its frequency. The dotted horizontal lines, labeled (1, 2, 3) in Fig. 2(a), identify typical driver frequency locations to be associated with the experimental responses illustrated in Fig. 1(a, b, c). To generate traveling ILMs one chooses the frequency of a linear mode in the band between the two arrows shown in Fig. 2(a). (This will become the carrier frequency of the traveling ILM.) Because of the small size of the system a well-defined frequency gap exists between that mode and the next higher frequency one as illustrated in Fig. 2(b) where the solid lines identify strongly driver active modes and the dashed lines weak driver active modes. The measured speed of each kind of traveling ILM is in good agreement with the slope of the dispersion curve at that particular driving frequency. These slopes are illustrated schematically in Fig. 2(a). The traveling ILMs shows some similarities and some differences with bright and dark solitons [15].

Numerical simulations for arrays of different sizes using the lumped element ball and nonlinear spring model described previously [14] support these findings. The open circles in Fig. 2(b) identify frequencies where locked smoothly traveling ILMs are found for three different system sizes: 50, 100, and 200 elements. Crosses correspond to frequency regions of instability, involving intermittent modes and complex traveling modes. The complexity comes about when some overtone of the ILM frequency couples to the linear phonon band. In general, smoothly traveling locked ILMs appear at frequencies above the unstable regions. Driving acceleration amplitudes are $2.5 \times 10^3$, $5 \times 10^3$ and $1 \times 10^4 m/s^2$, respectively. When the results are simple, like near mode at 112 kHz for 100 element case, the general sequence with increasing driver frequency is as follows: excite standing wave -> fluctuating standing wave -> complex traveling mode -> no response -> smoothly running ILM.

Figure 3 shows an averaged energy for the excitation (counting only the high amplitude cantilevers) as a function of driving frequency, for a single carrier mode at 112 kHz for N=100. The modulation instability starts at the frequency indicated by the left edge of the double-headed arrow. Until 112.5 kHz , a complex traveling mode, displayed in the inset (a), is observed. At frequencies above the energy gap the locked running ILM state appears as shown in insert (b). This localized feature is one side of a bi-stable state, the other side is a no-mode state, characteristic of the hysteresis response of a Duffing oscillator. As the linear modes become closely spaced with increasing system size, in addition to the overtone frequencies of the ILM coupling to the plane wave modes, the driven ILM from the N th carrier mode interacts with the driven complex mode of the N+1 th mode and becomes unstable.

To test the generality of these findings we have also numerically examined for limited sizes the dynamical properties of an easy plane antiferromagnetic system, the quasi 1-D $C_2H_5NH_3)_2CuCl_4$[9]. The equation of motion for a normalized spin $\vec{S}_n$ is

$$\frac{d}{dt}\vec{S}_n = -\gamma \vec{S}_n \times \vec{H}_n - \gamma \lambda \vec{S}_n \times (\vec{S}_n \times \vec{H}_n), \tag{1}$$

where $\vec{S}_n$ is a macroscopic spin averaged over ferromagnetic *ab* plane, $\gamma$, the gyromagnetic ratio and $\lambda$ is the Landau damping parameter. Here $\vec{H}_n$ is the magnetic field acting on the nth spin, with

$$\vec{H}_n = -2J(\vec{S}_{n-1} + \vec{S}_{n+1}) - 2\vec{\vec{A}}\vec{S}_n + H_{x0}\vec{e}_x \cos \omega t, \tag{2}$$

where $J$ is the nearest neighbor antiferromagnetic exchange constant, $\vec{\vec{A}}$, the anisotropic field tensor, and the last term the ac driving field along a specific transverse spin direction. The linear dispersion curve of the spin wave spectrum is shown in Fig. 4(a). At $k = 0$ the two magnetic active modes are linearly polarized in orthogonal directions. Since the nonlinearity of this dynamical equation is soft the frequency region between the two arrows shown is examined for bright traveling ILMs in limited systems. Shown in Fig. 4(b) are the spin wave modes of the lower branch that can be excited by the ac field polarized along the upper branch direction made active by the free boundary condition. (Using this polarization suppresses exciting the strong low frequency $k = 0$ mode.) The circles and x's again identify frequencies where either single traveling ILMs have or have not been generated. The insert in Fig. 4(a) shows such a locked traveling ILM at 1.74 GHz for the 100 spin lattice.

Our experiments demonstrate that micromechanical cantilever arrays of a few hundred elements can support locked ILMs of both the stationary and running type by simply tuning the driver frequency. Stationary locked ILMs, outside of the plane wave spectrum, are well known and can occur for any size discrete system. The new excitations are the locked, smoothly running acoustic ILMs that require the uniform driver frequency to occur in the frequency gap between neighboring plane wave modes. How the symmetry of the system is broken in the presence of a spatially uniform driver resulting in a single traveling ILM has not been resolved; however, practically these running localized excitations are a property of a small discrete system and they occur beyond the q-breather amplitude range. It should be mentioned that a theoretical study of a FPU 1-D lattice with a finite k driver at inband frequencies, but without the limited size restriction, has found only traveling modulated waves[16]. Finally since locked running ILMs depend only on discreteness, nonlinearity, the mode density, and the balancing of nonlinearity against the dispersion, the same dynamical properties can be expected to occur in a variety of small systems, including atomic lattice of nanoscale dimensions.

**Acknowledgments**

We dedicate this letter to the memory of Shozo Takeno, who pioneered in this field. MS would like to thank Kenichi Maruno for discussions about solitons. This work was supported in part by NSF DMR 0301035, by DOE DE-FG02-04ER46154 and by JSPS-Grant-in-Aid for Scientific Research (B)18340086.

**Figure captions**

**Figure 1.** Several kind of ILMs produced experimentally with the driver at different frequencies in a cantilever versus time plot. (a) Stationary bright ILM is recorded as a thick black horizontal band, (b) bright traveling ILM appears as a dark zigzag pattern, (c) dark traveling ILM zigzag pattern identified by the arrows. Where the dark ILM excitation travels the large amplitude standing wave pattern vanishes. The bright regions at its boundaries are due to the overlap of the incident and reflected hole amplitudes at those locations. Driving frequencies: (a) 137.8 kHz, (b) 110.086 kHz, and (c) 81.26 kHz.

**Figure 2(a).** Linear and nonlinear dispersion map of the di-element cantilever array showing various locked ILM positions. Open circles identify the linear modes. The three horizontal dashed lines identify characteristic driving frequencies. (1) Solid diamonds, illustration of a Fourier transform (FT) of a locked, stationary ILM generated above the top of the upper branch, (2) open squares, FT of a locked traveling bright ILM and (3) open squares, FT of a locked traveling dark ILM. Solid circles indicate carrier modes of traveling ILMs. These locked ILMs occur where the dispersion balances the nonlinearity and map as groups of points on straight lines. Because of the hard nonlinearity the bright ILM, (2), appears above the linear dispersion curve while the dark ILM, (3), appears below the nonlinear dispersion curve (dotted curve that is shifted up by the background excitation of the dark ILM). The hole feels a soft nonlinearity in the strongly excited background. The two arrows indicate the frequency region examined numerically.

Figure 2(b). Numerical simulation results for the dependence of traveling locked bright ILMs on system size. Solid and dotted horizontal lines show strongly and weakly active vibrational modes, respectively. Open circles identify frequencies where locked smoothly traveling ILMs are found. Crosses correspond to frequency regions of instability, involving intermittent modes and complex traveling modes. The smoothly traveling locked ILMs appear at frequencies above these unstable regions.

Figure 3. Energy of cantilever per site averaged over the large amplitude cantilevers as a function of driver frequency for N=100. The complex traveling mode pattern (insert a) occurs in the frequency region indicated by the two heaeded arrow. The smooth traveling ILM (insert b) is observed for frequencies between 112.54 to 113.6 kHz. The frequencies of inserts (a) and (b) are 112.45 kHz and 112.6 kHz, respectively. The time window of each inserts is 4ms. The spikes occur at instabilities within the averaging 4000 period time duration.

Figure 4(a). Linear dispersion curve of the quasi 1-D antiferromagnet $(C_2H_5NH_3)_2CuCl_4$. The two magnetically active $k = 0$ modes are linearly polarized in orthogonal directions. Because of the soft nonlinearity of magnetic systems the lower frequency region between the two arrows will be probed numerically for locked traveling bright magnetic ILMs. Insert: Spin site versus time showing a single locked traveling magnetic ILM at 1.74 GHz for the 100 spin case. The insert time window is 57 ns.

Figure 4(b). Numerical simulation results for the dependence of traveling locked ILMs on system size. Horizontal lines identify spin wave modes of the lower branch that can be excited by the ac field polarized along the upper branch $k = 0$ direction due to the finite lattice size with free boundary conditions. Crosses indicate frequencies for the complex traveling modes and open circles identify smoothly traveling mode locations. The frequency of the smoothly running mode is always lower than the complex mode, when they stem from the same carrier mode.

Figures

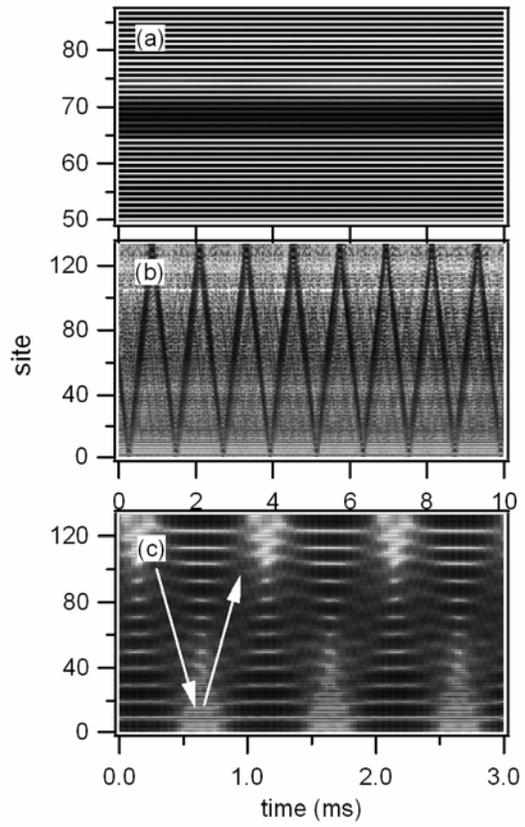

Fig. 1.

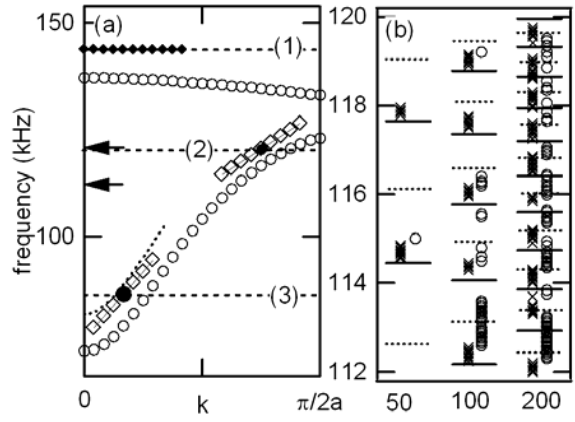

Fig. 2.

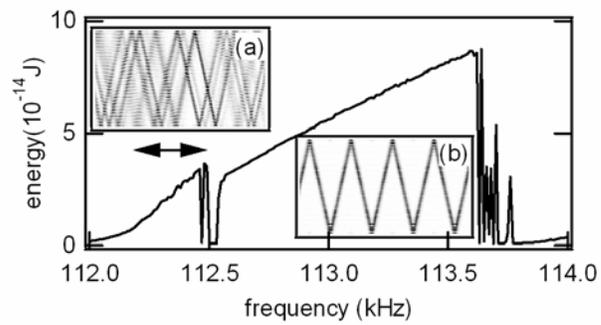

Fig. 3.

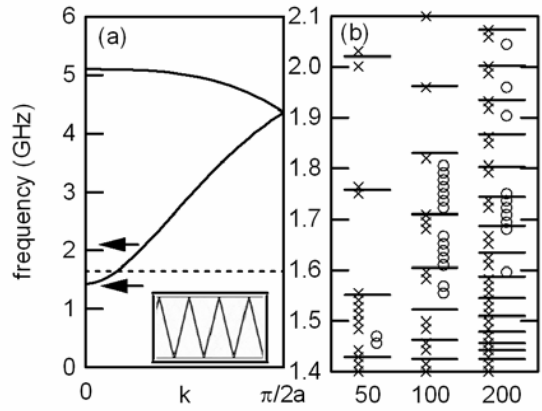

Fig. 4.